\documentclass[aps,prb,twocolumn,amsmath,amssymb]{revtex4-1}
\usepackage{graphicx,graphics,subfigure}% Include figure files
\usepackage{epstopdf}
\usepackage{dcolumn,mathtools}% Align table columns on decimal point
\usepackage{color,bm}% bold math

\begin{document}
\title{\textit{Ab initio} thermodynamics of intrinsic oxygen vacancies in ceria}
\author{Chirranjeevi Balaji Gopal$^1$ and Axel van de Walle$^2$}
\affiliation{$^1$Department of Applied Physics and Materials Science, California Institute of Technology, Pasadena, California 91125, USA, $^2$School of Engineering, Brown University, Providence, Rhode Island 02912, USA}
\date{\today}

\begin{abstract}
Nonstoichiometric ceria(CeO$_{2-\delta}$) is a candidate reaction medium to facilitate two step water splitting cycles and generate hydrogen. Improving upon its thermodynamic suitability through doping requires an understanding of its vacancy thermodynamics. Using  density functional theory(DFT) calculations and a cluster expansion based  Monte Carlo simulations, we have  studied the  high temperature thermodynamics of  intrinsic oxygen vacancies in ceria. The DFT+$U$ approach  was used to get the ground state energies of various vacancy configurations in ceria,  which were subsequently 
fit to a cluster expansion Hamiltonian to efficiently model the configurational dependence of  energy. The effect  of lattice vibrations was incorporated  through a temperature dependent cluster expansion. Lattice Monte Carlo simulations  using the cluster expansion Hamiltonian  were able to detect the miscibility gap in the phase diagram of ceria. The inclusion of vibrational and electronic entropy effects made the agreement with experiments quantitative. The deviation from an ideal solution model was quantified by calculating as a function of nonstoichiometry, a) the solid state entropy from Monte Carlo simulations and b) Warren-Cowley short range order parameters of various pair clusters. 
\end{abstract}

\maketitle

% body of paper here - Use proper section commands
% References should be done using the \cite, \ref, and \label commands
\let\clearpage\relax
\section{\label{sec:intro}Introduction}
Ceria(CeO$_2$) and ceria based fluorite structure oxides are among the best performing solid oxide fuel cell electrolytes and have historically been used as automotive exhaust catalysts to reduce noxious gases. Due to its structural similarity to urania and thoria, ceria is also of interest to the nuclear fuels community to investigate radiation damage to fuel rods in nuclear reactors. Recently, Chueh et al.\cite{Chueh2010,Chueh2010b} demonstrated a two step solar driven 
 thermochemical cycle based on ceria to split water and produce hydrogen as follows. 
\begin{align}\label{eq:reaction}
&\mathrm{CeO}_{2-\delta_L} \xrightarrow{T_H} \mathrm{CeO}_{2-\delta_H} + \frac{\delta_H-\delta_L}{2}\mathrm{O}_2\\
&\mathrm{CeO}_{2-\delta_H} + (\delta_H - \delta_L) \mathrm{H}_2\mathrm{O} \xrightarrow{T_L} \mathrm{CeO}_{2-\delta_L} + (\delta_H -\delta_L) \mathrm{H}_2 
\end{align}
$\delta_L$ and $\delta_H$ respectively denote the oxygen nonstoichiometry at the low temperature (T$_L$) and  high temperature (T$_H$) steps. The high temperature  step of the redox cycle  involves release of oxygen from ceria, forming oxygen vacancies(henceforth referred to as vacancies) in the lattice. At the low temperature  step, ceria reacts with water, oxidizing itself and  liberating H$_2$ which can be used to generate electricity through a fuel cell. From a purely thermodynamics perspective, these applications stem from the remarkable ability of ceria to display significant oxygen nonstoichiometry without changing its crystal structure. The oxygen vacancy formation is facilitated by the ability of Ce to exist in two oxidation states: Ce$^{3+}$ and Ce$^{4+}$. A general thermodynamic framework based on DFT+\textit{U} calculations to assess the suitability of an oxide for two-step water splitting cycles was proposed by Meredig and Wolverton\cite{Meredig2009}.

Experimental work on undoped ceria has been actively pursued for a long time, excellent reviews of which have been published by Inaba et al.\cite{Inaba1996} and Mogensen et al.\cite{Mogensen2000}. Thermodynamic data including nonstoichiometry $\delta$ as a function of temperature (T) and  partial pressure of oxygen ($p_{O_2}$) and the enthalpy and entropy of reduction reaction has been measured extensively\cite{Panlener1975}. On the computational side, significant research has been devoted to understanding the electronic structure of nonstoichiometric ceria from first principles using density functional theory(DFT)\cite{Andersson2007,Castleton2007}  with the standard local density approximation(LDA) and generalized gradient approximation(GGA) exchange correlation functionals. It is well known that the 4f electrons in CeO$_2$ need to be treated as valence states to accurately reproduce the experimentally observed properties\cite{Skorodumova2001}. Using the conventional LDA and GGA exchange correlation functionals in ceria leads to the self interaction error and a consequent  failure to reproduce the insulating character of defect-free ceria. This necessitates adding a Hubbard  potential($U$) correction for the Ce `4f' electrons to generate the experimentally observed band gap\cite{Loschen2007}.  Hybrid functionals yield improvement in the electronic picture of ceria, but  do not significantly change the energetics of vacancy formation\cite{Silva2007}. The effects of transition and rare earth metal dopants\cite{Gupta2010,Scanlon2011,Zyang2008} : both aliovalent\cite{Andersson2006,Andersson2007a} and isovalent\cite{Woo2006,Chen2010} on oxygen vacancy formation  have been investigated using a supercell approach. The oxygen storage capacity is correlated to the structural relaxation brought about by dopants with smaller ionic radius than Ce$^{4+}$ and the electrostatic effects. Activation energies for potential vacancy migration pathways have been computed from first principles\cite{Dholabhai2010,Ismail2011,Dholabhai2010b} to understand the mechanisms of defect migration at atomistic scale. With the development of density functional perturbation theory, lattice dynamical properties, Born effective charges and phonon density of states have been calculated and these are found to agree well with experimental data\cite{Gurel2006,Dutta2010}.

 However, \textit{ab initio} calculations are currently intractable for system sizes greater than a few hundred atoms. To obtain finite temperature bulk properties, it is necessary to move beyond the realm of DFT calculations to statistical thermodynamics of larger system sizes( \textgreater$10^3$ atoms ). While literature abounds with experimental investigations of the properties of ceria at high temperatures, the computational work described above has been primarily limited to studying the electronic structure and defect formation energies of ceria at absolute zero without attempting to obtain thermodynamic properties at higher temperatures, which is where its most interesting applications and properties emerge.
 
Our computational study aims to isolate and focus on the thermodynamics of intrinsic oxygen vacancies in ceria relevant to thermochemical cycling.  While vacancies are  central to the thermochemical cycling process, the thermodynamic driving force is governed by the change in $\delta$ (i.e. $\Delta\delta$) between $T_H$ and $T_L$. For repeatability of the redox cycle, it is necessary to operate in the regime of single phase nonstoichiometric ceria. Furthermore, at such high values of $\delta$, defect interactions become considerable and can negatively impact the entropic  driving force for the reduction of ceria. Understanding the nature of vacancy interactions as a function of temperature and concentration in ceria will therefore be instrumental in motivating dopants to improve its suitability. Aside from its immediate relevance to thermochemical fuel production, we chose ceria as our model system since its phase diagram and vacancy thermodynamics are established experimentally.  This would help test the accuracy of the computational thermodynamics approach to studying nonstoichiometric oxides, and  extend it to screen dopants and predict properties of doped ceria, for which literature is not as extensive.

Modeling of ceria presents a unique challenge: it requires the ability to capture electron-localization and associated electronic entropy effects. A step in this direction was made for the Li$_x$FePO$_4$ system\cite{Zhou2006} by showing that including electronic entropy via a cluster expansion approach yields a phase diagram whose topology is in qualitative agreement with experiments. We build upon this effort and seek to conclusively demonstrate the thermodynamic validity of such an approach, by verifying that the inclusion of both electronic and vibrational entropies results in excellent quantitative agreement with experiments for not only the phase diagram, but other thermodynamic quantities as well, such as the entropy of reduction and short-range order.

The paper is organized as follows. The methodology of studying phase equilibria from first principles as applied to ceria will be discussed in Section ~\ref{sec:method}. Results and discussion (section ~\ref{sec:results}) will be broken down into subsections to focus independently on electronic structure calculations,  cluster expansion and free energy integration. Wherever applicable, the intricacies of dealing with intrinsic oxygen vacancies in ceria will be emphasized. 

\section{\label{sec:method}Methodology}
%AVDW: ref added:
The standard first principles approach to computing phase equilibria has been detailed previously.\cite{Ruban2008,Ceder2000,Walle2002a} We present a brief overview here to highlight salient features of this approach relevant to nonstoichiometric oxides. Phase equilibria studies from first principles integrate both rigorous first principles calculations over selected small structures and large scale statistical ensemble based methods. The partition function, which contains all the thermodynamic information for a system, can be coarse grained\cite{} over a hierarchy of degrees of freedom as described in Equation~\ref{eq:Z}.
%AVDW
\begin{equation}\label{eq:Z}
Z=\sum_{\sigma\in L}\sum_{\nu\in\sigma}\exp\left[-\beta E(\sigma,\nu)\right]
\end{equation}    
where $\nu$,$\sigma$ are respectively the vibrational and configurational states of the system constrained to a lattice L. Here, the configurational states include both genuine configuration variables (the presence of oxygen vacancies) and electronic state information (the location of the Ce$^{3+}$ ions. The energy $E(\sigma,\nu)$ is obtained by performing quantum mechanical calculations for a fixed composition in the lattice, followed by ionic relaxation. Vibrational frequencies of the system are then calculated for small displacements away from the relaxed ground state at 0 K. The cluster expansion then parametrizes  this information in terms of larger structural units and enables estimation of energies of  cells and compositions inaccessible through first principles calculations in a fast and efficient manner. Finally, the thermodynamic integration procedure, with the aid of Monte Carlo simulations incorporates the effect of compositional fluctuations and temperature  on the properties of the system and is used to derive other thermodynamic quantities.
 
 \subsection{First principles calculations}
                                                                
 Electronic structure calculations were performed using the Vienna \textbf{\textit{ab initio}} simulation package (VASP),  a plane wave pseudopotential based DFT code\cite{Kresse1996}. GGA exchange correlation functional using the PAW (projector augmented wave) method\cite{Blochl1994} was employed. To correct for the strong on-site coulombic interaction of the Ce 4f electrons, we adopted the rotationally invariant GGA+\textit{U} formalism introduced by Dudarev et al\cite{Dudarev1998}.  The Hubbard potential term `\textit{U}' penalizes partial occupancy of the f states and opens up a band gap. The value of U is typically set by fitting to  experimentally established band gaps, or quantities such as lattice constant and bulk modulus. For nonstoichiometric ceria, based on previous LDA+\textit{U} and GGA+\textit{U} studies\cite{Castleton2007,Loschen2007} of oxygen vacancy formation energies and electron localization on Ce$^{3+}, $ \textit{U}=5 for GGA+\textit{U} and \textit{U}=6 for LDA+\textit{U} have been proposed as optimal. Spin polarized GGA+\textit{U} calculations in this work were all performed using \textit{U}=5.
 
Ceria has a fluorite structure, with oxygen atoms occupying the tetrahedral voids of an FCC lattice of Ce. Defect calculations were performed on 2x2x2 supercells of ceria (96 atoms), using a 2x2x2 \textit{k}-point grid.  Electronic relaxations were performed until the total energy difference was  less than $10^{-4}$ eV while ionic relaxations  were carried out until residual forces  less than $0.02$ eV/$\AA$ were achieved. Formation of vacancies in ceria leads to expansion of the lattice resulting from increased coulombic repulsion  between the Ce$^{4+}$ ions and larger ionic radius of Ce$^{3+}$. We account for this by performing multiple constant volume relaxations at distinct volumes for each structure. The energy benefit accrued from the volume relaxation is significant, and if overlooked, could lead to erroneous  energies predicted with the cluster expansion later on. In all, 36 different configurations of vacancies were studied with compositions ranging from CeO$_2$ to CeO$_{1.75}$.
%The MAGMOM tag in VASP was used to ascertain the difference between $Ce^{4+}$ and $Ce^{3+}$ (with a 4f electron).
 
First principles lattice dynamics calculations were performed to incorporate vibrational effects on phase stability. We used the `small displacement' finite differences method as implemented in VASP 5.2 to compute the Hessian matrix for the structures. Displacements of 0.015 $\AA$  away from the equilibrium relaxed positions were employed. For higher vacancy concentrations, we included structures with both pseudo-randomly dispersed and clustered arrangements to span the configurational dependence of the vibrational frequencies. The force constants output by VASP  were used to obtain the  dynamical matrix at other q-points and calculate out the phonon density of states (DOS) using PHONOPY\cite{phonopy}. Gaussian smearing and an 8x8x8 q-point mesh were employed for the DOS calculation. The vibrational free energy (F$_\mathrm{{Vib}}$) and  entropy (S$_\mathrm{{Vib}}$) were evaluated under the harmonic approximation as 

\begin{equation}
\frac{F_{Vib}(T)}{N}=\frac{E^*}{N} +
 k_BT\int_0^\infty \ln\left(2 \sinh\left(\frac{h\nu}{2k_BT}\right)\right)g(\nu)d\nu 
 \end{equation}
 \begin{equation}
\frac{S_{Vib}(T)}{N} = \left(\frac{\partial F_{Vib}/N}{\partial T}\right)_V
\end{equation} 
 where N is the total number of atoms in the system, $\nu$ is frequency of phonon mode, g($\nu$) is the phonon density of states.
  
 \subsection{Cluster expansion : configurational degrees of freedom}
 The cluster expansion (CE) Hamiltonian treats configurational disorder by decomposing the energy of a lattice into a basis of clusters ( points, pairs, triplets etc.) of lattice sites. Each cluster is a polynomial in occupation variables and has an associated effective cluster interaction (ECI), `J' in equation \ref{eq:CE}. The ECIs are obtained by fitting to the database of \textit{ab initio} energies. 
\begin{equation}\label{eq:CE}
E(\sigma) = J_0 + \sum_i J_i\sigma_i +\sum_{i\neq j}J_{ij}\sigma_i\sigma_j + ...
\end{equation}
Vacancies are treated as independent species, so any site in the anion sub-lattice can be occupied by an oxygen ion or vacancy. Additionally, 
we explicitly  treat charge state disorder in the cation sub-lattice  resulting from Ce$^{4+}$/Ce$^{3+}$ ( configurational electronic entropy ). In order to describe the energetics of this system with two interacting sub-lattices, we use a multicomponent multilattice CE
formalism\cite{vandeWalle2009,tepesch1995} that works in the product basis of cluster functions defined on each sublattice. Despite the presence of 4 distinct species (O, Vac, Ce$^{3+}$, Ce$^{4+}$), constraints of site and charge balance  (2[Ce$^{3+}$] = [Vac]) render the system essentially  pseudo-binary. Our cluster expansion fit was obtained using the mmaps code in the alloy theoretic automated toolkit (ATAT)\cite{Walle2002a,Vandewalle2002}.

The knowledge of ECIs provides a computationally inexpensive and efficient means to compute the energy of a large system on the fly during  Monte Carlo simulations, circumventing the need for time consuming \textit{ab initio} calculations. As such, the CE fit to \textit{ab initio} energies is independent of temperature. By cluster expanding phonon free energies in the basis of clusters fit to configurational energies, the ECIs, and consequently the MC data can be made to include vibrational effects.

\subsection{Lattice Monte Carlo simulations : thermodynamic properties}
The fundamental external variables of interest to the thermochemical cycling of ceria are temperature (T) and oxygen partial pressure ($p_{O_2}$). Its properties are strongly dependent on the oxygen nonstoichiometry, which is uniquely set for a given (T, $p_{O_2}$). Thus, ceria lends itself to be conveniently studied by  semi grand canonical Monte Carlo simulations, treating temperature and chemical potential as external variables. A semi-grand canonical ensemble fixes the total number of lattice sites, letting the concentration of individual species fluctuate in response to an an applied temperature or  chemical potential change. Consideration of macroscopic charge-neutrality leaves us with  one independent chemical potential $\mu$ written as :  
\begin{equation}
\mu = (\mu_{Vac}-\mu_{O^{2-}}) + 2(\mu_{Ce^{3+}}-\mu_{Ce^{4+}})
\end{equation}
where $\mu_{Vac},\mu_{O^{2-}},\mu_{Ce^{3+}}$ and $\mu_{Ce^{4+}}$ denote the chemical potentials of the individual species which are externally imposed. $\mu$ can be described as the free energy cost associated with swapping a pair of Ce$^{4+}$ and O$^{2-}$ with a pair of Ce$^{3+}$ and $Vac$. 
 
The Grand Potential $\Phi(\mu,T)$ with respect to a reference can be obtained by thermodynamic integration along a fixed  T or fixed $\mu$ path
\begin{equation}\label{eq:TDintegration1}
\Phi(T_0,\mu) = \Phi(T_0,\mu_0) -\int^{\mu}_{\mu_0}\left<N(T_0,\mu)\right>d\mu
%AVDW N is not the right variable!
\end{equation}

\begin{equation}\label{eq:TDintegration2}
\frac{\Phi(T,\mu_0)}{k_BT} = \frac{\Phi(T_0,\mu_0)}{k_BT_0} +
%\int^{T}_{T_0}(\left<E(T_0,\mu)\right> - \boldsymbol{\mu\left<N(T,\mu_0)\right>})d\beta
\int^{T}_{T_0}(\left<E(T,\mu_0)\right> - \mu\left<N(T,\mu_0)\right>)d\beta
\end{equation}

 $\langle E \rangle$ is the thermodynamically averaged energy, $\left<N\right>$ is the thermodynamically averaged concentration, $\mu_0$ and T$_0$ are the reference chemical potential and temperature,  and $\beta = 1/k_BT$. The low temperature and high temperature  expansions to obtain the reference points for integration are not central to this paper and can be found elsewhere\cite{Walle2002}. Our MC runs were performed on 5184 atom cells (larger sizes were attempted and found to not affect the results) using the memc2 code  from ATAT\cite{vandeWalle2009}.
 Temperature steps of 40 K were used for the MC runs, with 2000 equilibrium passes and 1000 averaging passes at each  ($T,p_{O_2}$). Simultaneous spin flips were used to maintain charge balance and were constrained to occur within two unit cell distances of each other.

It should be noted that the constraint of charge balance alone is not sufficient to guarantee that the system will never undergo a phase separation into multiple spurious ground states that are locally non charge balanced although the overall simulation cell is. This can occur when the cluster expansion is only fitted to charge-balanced structures, thus providing little guarantee that the extrapolated energy of non-charge-balanced structures is physically meaningful.
We avoided this problem by an iterative procedure. Starting with a cluster expansion fitted to charge-balanced structure only, we monitored the simulation for evidence of phase separation into non-charge-balanced structures. Whenever this was observed, the energy of the structure was calculated from first principles and included in the cluster expansion to reduce extrapolation errors from the fit. Since \textit{ab initio} calculations imposing periodic boundary conditions necessarily enforce charge-neutrality, we used a neutralizing background charge to estimate the energy of  non-charge-balanced structures. This ansatz is justified whenever the resulting calculated energies are sufficiently high, so that these configurations are very rarely sampled in equilibrium.)

\section{\label{sec:results}Results and Discussion}
\subsection{\label{subsec:abinitio}First principles}
Using GGA+\textit{U} calculations, the equilibrium lattice constant for stoichiometric ceria was found to be 5.48$\AA$. 
The formation of an oxygen vacancy in a 2x2x2 supercell is accompanied by the localization of two electrons onto the 4f states of two  Ce atoms with anti-symmetric spins. 
%In the lowest energy configuration, the two electrons localize with anti-symmetric spins on 2 of the 4 tetrahedrally coordinated Ce atoms. 
True ground state convergence in other vacancy structures was tested by assigning different starting spin configurations to the Ce atoms (up or down spin, and their locations) and looking for the lowest energy structure. The  2p-5d and 2p-4f energy gaps were 5.3 eV and  2.5 eV respectively. The formation energy for an oxygen vacancy in a 2x2x2 supercell was calculated to be 3.2 eV, agreeing with previous GGA+\textit{U} studies\cite{Andersson2007}. We included up to 8 vacancies in a 2x2x2 supercell (corresponding to CeO$_{1.75}$) in different configurations. Such  defect concentrations, while high, have  been shown to exist in ceria under appropriate $(T, p_{O_2})$ and are of interest to us. The lattice expansion associated with vacancy formation was evaluated through multiple constant volume relaxations of each configuration and found to be around 2\% for $CeO_{1.75}$ (8 vacancies in a 2x2x2 supercell). 

\begin{figure} 
\includegraphics[scale=0.75]{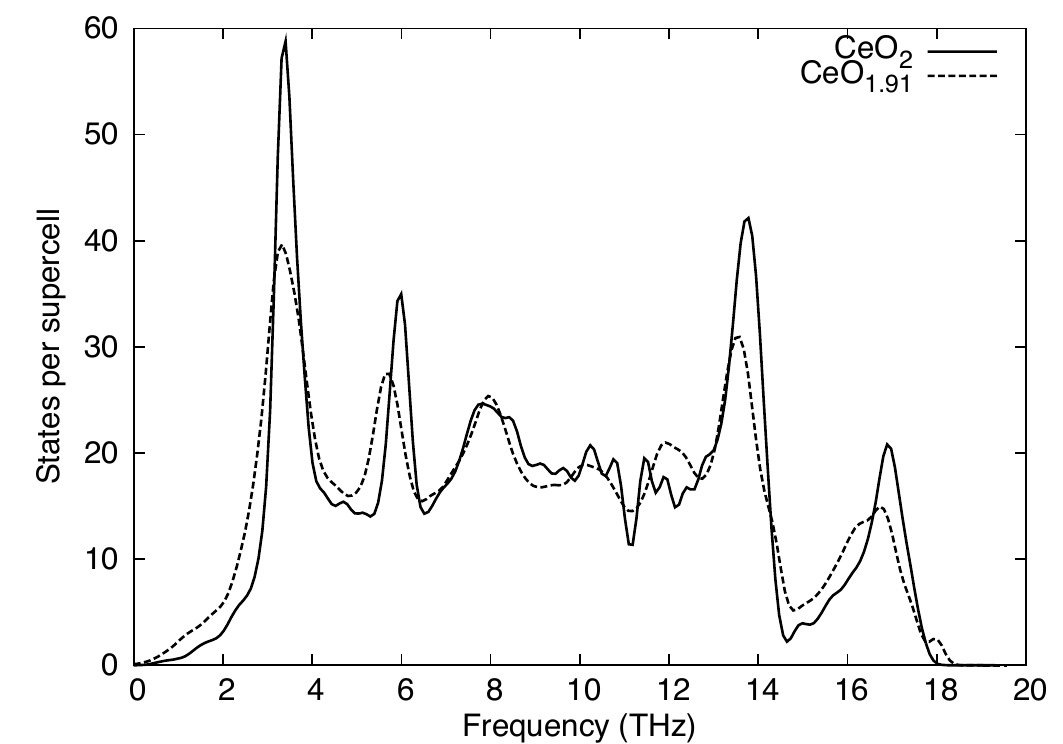}% Here is how to import EPS art
\caption{\label{fig:vdos} Phonon density of states from first principles calculated under the harmonic approximation. Oxygen vacancies lead to softening of vibrational modes in CeO$_{1.91}$ (3 vacancies in a 2x2x2 supercell) compared to CeO$_2$.}
\end{figure}

Phonon DOS computed  under the harmonic approximation are plotted in Fig.~\ref{fig:vdos}. Vacancies are clearly stabilized by vibrations as evidenced by the softening of modes in CeO$_{1.91}$ (3 vacancies in a 2x2x2 supercell) compared to CeO$_2$. Further, for a given concentration, clustered vacancies had stiffer phonon modes than vacancies dispersed over the supercell.

\subsection{\label{subsec:CE}Cluster Expansion}
\begin{figure}[b]
\includegraphics[scale=0.47]{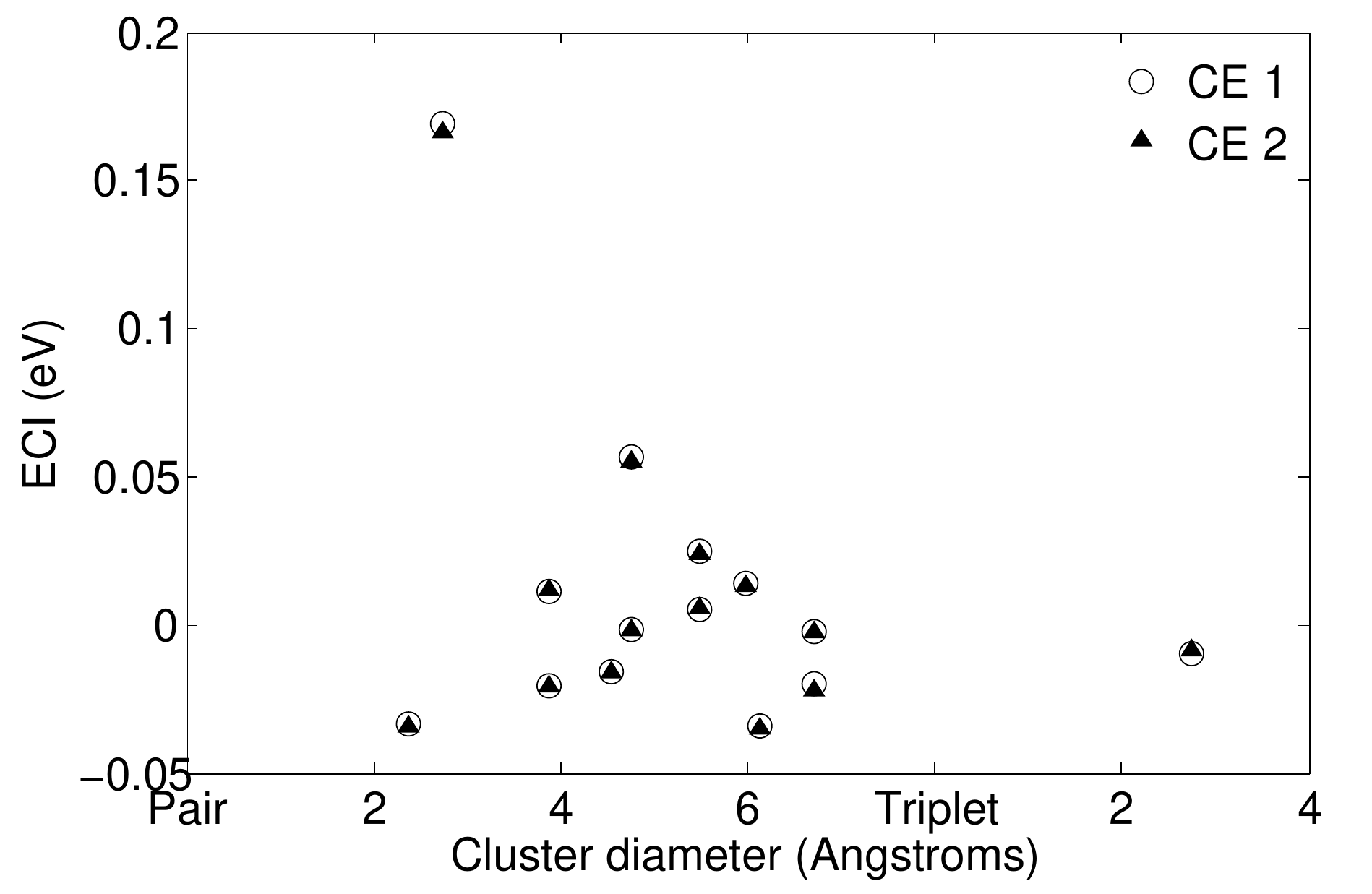}
\caption{ECI vs cluster diameter for a cluster expansion fit to the as calculated first principles energies (squares, denoted by CE 1). Subtracting the electrostatic energy prior to fitting and later adding it back (as an energetic correction) does not change the variation of ECIs with cluster diameter (filled triangles, CE 2). This indicates that the electrostatic interactions are captured by the cluster expansion and do not need an explicit treatment.}
\label{fig:ecis}
\end{figure}

%AVDW:
The optimal cluster expansion fit to the calculated \textit{ab initio} energies (CE 1) of the 36 relaxed geometries comprised 13 pairs and 1 triplet clusters apart from the null and 2 point clusters.  Considering that ceria is an ionic compound, it is important to ascertain if the long range electrostatic interactions are adequately described by a short-range cluster expansion. To ascertain this, we fit another cluster expansion (CE 2) to the \textit{ab initio} energies after subtracting out the coulombic energy term (computed through an Ewald summation).  The latter was then cluster expanded in the basis of clusters identified by the fit, and added as an energetic correction to the ECIs. The results are illustrated by Fig.~\ref{fig:ecis}. The pair and triplet cluster ECIs of CE 2 show identical decay characteristics with cluster diameter as CE 1, indicating that electrostatic interactions are captured well by the cluster expansion. The cross-validation score\cite{Walle2002a}, which provides a measure of the predictive power  of a cluster expansion fit, was close to 0.003 eV for both CE 1 and CE 2. In view of these results, we justify using CE 1 for further work, given its higher computational efficiency.
  
The effect of temperature on ECIs was incorporated  by cluster expanding the vibrational free energies in the basis of clusters of CE 1. This introduced a temperature dependent correction to 7 clusters (determined via cross-validation) out of a total of 17. 
%%%%%%%%%%%%%%%%%%%%%%%%%%%%%%%%%%%%%%%%%%%
\subsection{\label{subsec:MC}Monte Carlo Simulations}
The Ce-O phase diagram in the composition range of Ce$_2$O$_3$ - CeO$_2$ has been determined by experiments. The low temperature portion of the phase diagram is complicated by the many vacancy ordered phases of  composition Ce$_n$O$_{2n-m}$, and is not of primary concern to this work. Of interest to us is the high temperature phase diagram and the ability of first principles calculations to predict the miscibility gap and vacancy thermodynamics in single phase non stoichiometric ceria. 

\begin{figure}[b]
\hspace*{-0.3cm}\includegraphics[scale=.36]{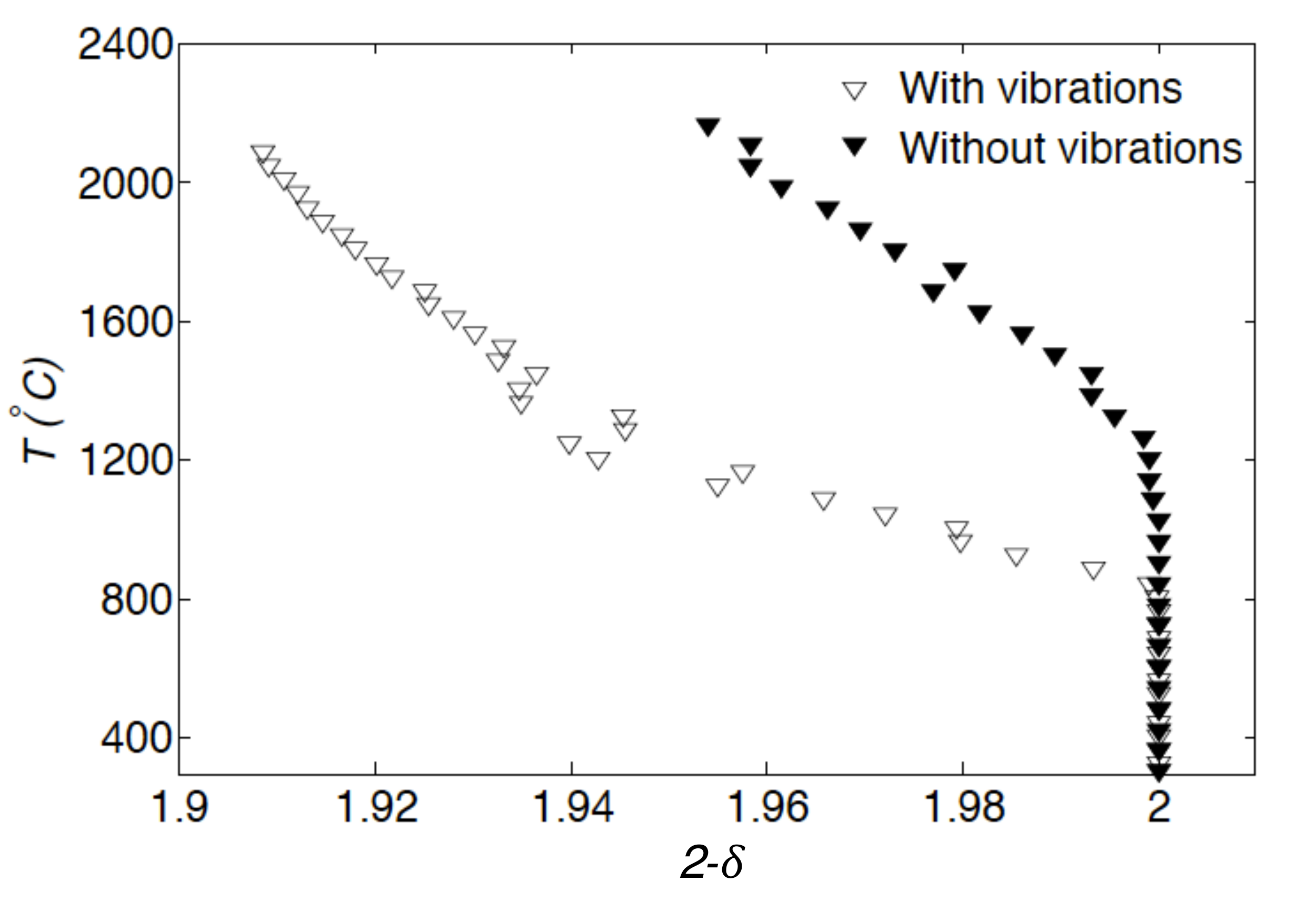}% Here is how to import EPS art
\caption{ A constant $\mu$ trajectory MC simulation starting from the low temperature ground state (CeO$_2$) illustrating  vibrational effects on the phase diagram and vacancy concentrations.}
\label{fig:constmu}
\end{figure}

A temperature-composition plot along a constant $\mu$ trajectory starting from 0 K is shown in Fig.~\ref{fig:constmu}. Defect free $CeO_2$, the starting ground state (GS), is stable up to  1300$^\circ$C before oxygen vacancies start to form. With the inclusion of vibrational effects, the onset of vacancies happens at a much lower temperature of 900$^\circ$C, all other variables being the same. Most \textit{ab initio} phase diagram studies have in past neglected the non-configurational contributions to defect formation entropy, which can substantially affect phase stability.

\begin{figure}
\includegraphics[scale=.36]{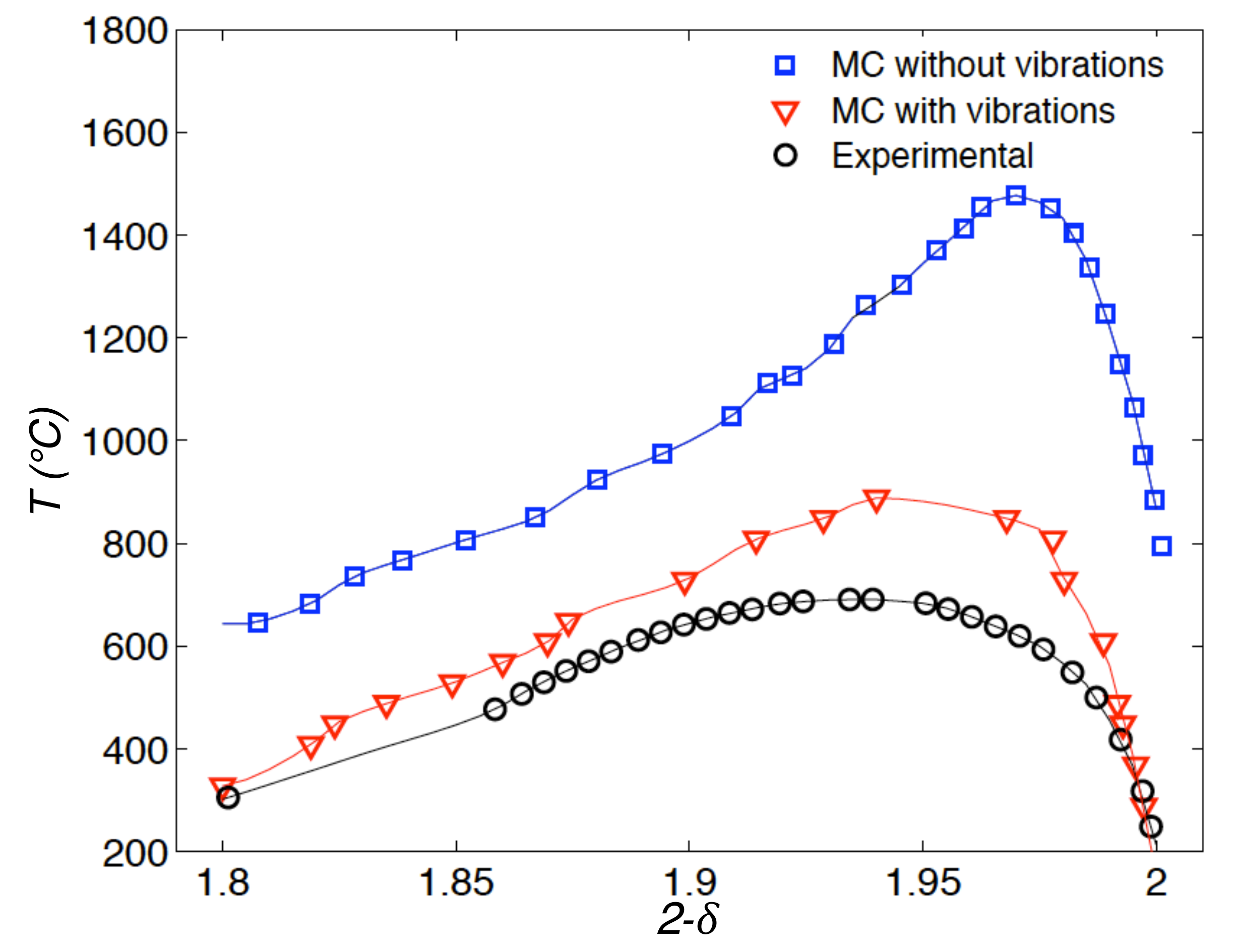}
\caption{(Color online) Miscibility gap in ceria calculated from Monte Carlo simulations. The pronounced effect of vibrations is visible from the suppression of the miscibility gap to lower temperatures and  enhanced vacancy solubility in nonstoichiometric ceria. The experimental phase diagram\cite{Ricken1984} has been overlaid for comparison. }
\label{fig:MG}
\end{figure}

In order to accurately predict the miscibility gap, it is necessary to perform simulations from two different starting points for each $\mu$ -- a heating simulation from the low temperature ordered ground state, and a cooling simulation from the high temperature disordered phase. At a given temperature, the discontinuity in  concentration when the grand potentials from the two runs are equated leads to the miscibility gap. The phase diagram in the composition range of  CeO$_{1.8}$ to CeO$_2$ obtained using this approach is shown in Fig.~\ref{fig:MG}. A miscibility gap shows up  even in the absence of vibrations, but the temperature scale is off by nearly a factor of two compared to experiments. The solubility limit of vacancies in ceria is underestimated and the miscibility gap is shown to persist up to temperatures as high as 1500$^\circ$C. The  temperature dependent cluster expansion provides a closer agreement : the miscibility gap closes at ~800$^\circ$C (690$^\circ$C in experiments) and  single phase ceria is shown to be stable at higher oxygen nonstoichiometry at any given temperature.

The cluster expansion technique was principally intended to model configurational disorder in alloys in which interatomic interactions tend to be much simpler than in insulators or semiconductors.  Studies have however expanded its domain of applications to describe the energetics of Li intercalation in battery materials and model charge state disorder through a localized electronic entropy term\cite{Zhou2006}, equilibrium composition profile across interfaces of doped ceria superlattices\cite{Walle2007}. This is the first study of high temperature phase diagram of ceria  from first principles.  That it captures the thermodynamics and detects a miscibility gap in a correlated electron system is in itself a significant result; the quantitative agreement with experiment upon including the effect of vibrations and electronic entropy shows even more promise.

\begin{figure}
\hspace*{-0.5cm}\includegraphics[scale=0.37]{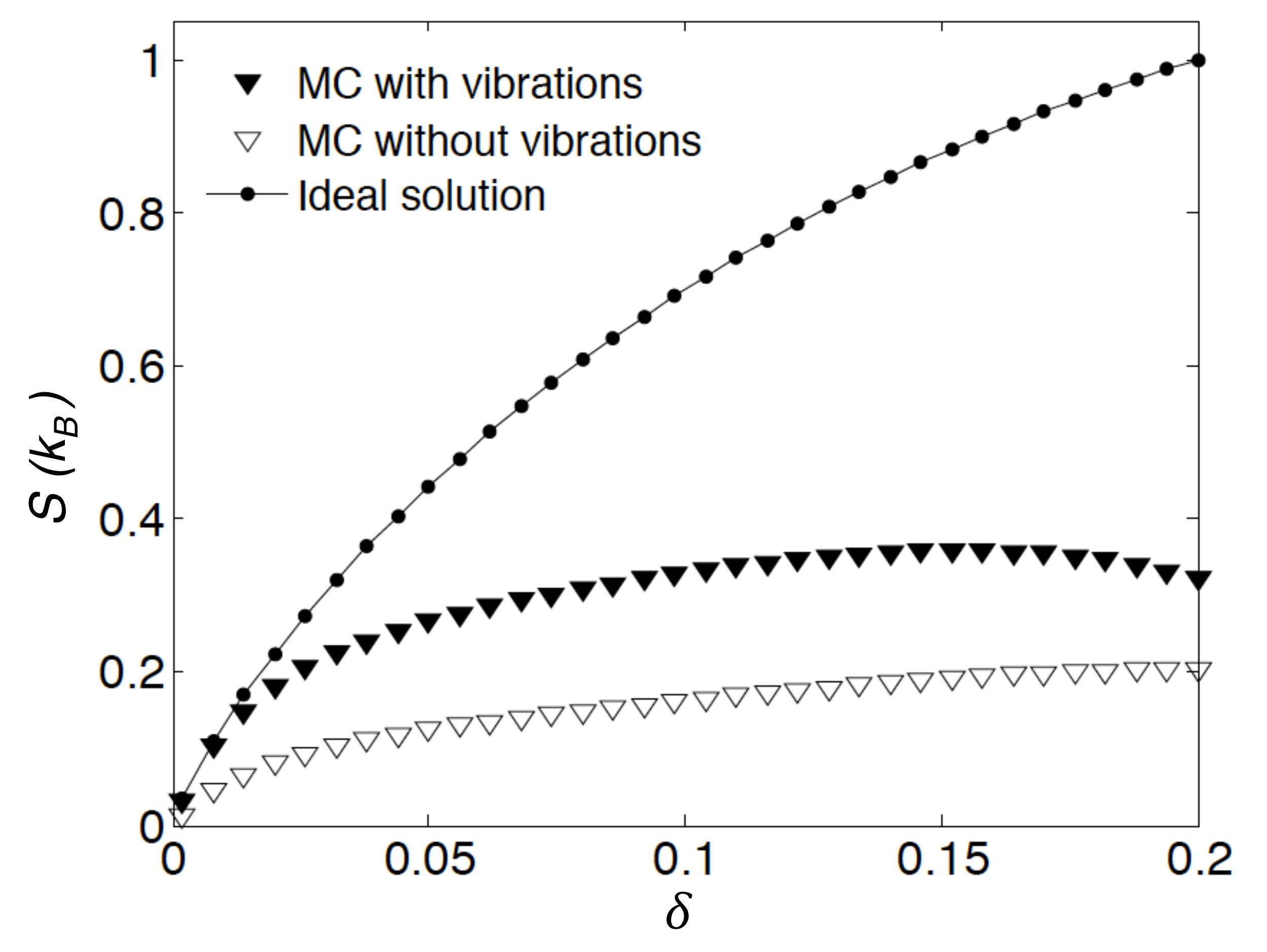}
\caption{Entropy of CeO$_{2-\delta}$ referenced to $CeO_{2}$ computed from MC simulations at 1480 K. The deviation from ideal solution behavior (configurational disorder in Ce$^{3+}$/Ce$^{4+}$ and O$^{2-}$/Vac lattice) becomes apparent even at low $\delta$. Vibrations provide entropic benefit to forming vacancies, but the marked non-ideal behavior prevails.}
\label{fig:entropydeviation}
\end{figure}

\begin{figure}
\hspace*{-0.5cm}\includegraphics[scale=0.38]{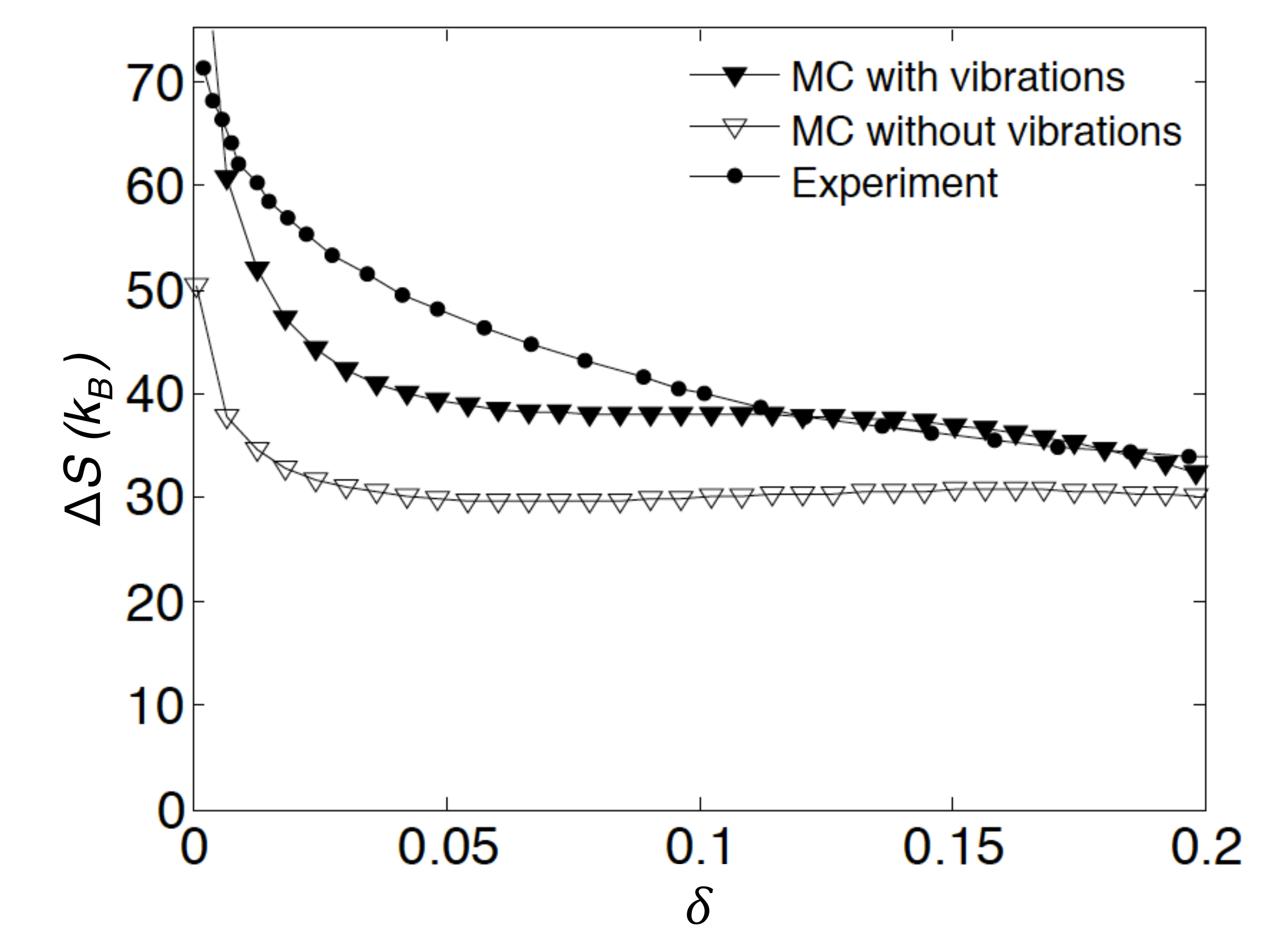}
\caption{Entropy change associated with forming an oxygen vacancy (see Eqn.~\ref{eq:reaction2}) as a function of nonstoichiometry compared with experiment\cite{Panlener1975} }
\label{fig:entropyreduction}
\end{figure}

The phase diagram helps identify thermodynamically stable phases at equilibrium, but even in the region of single phase nonstoichiometric ceria, significant short range  order can persist leading to deviation of bulk thermodynamic properties from that of an ideal solution. Indeed, this is the case and ceria deviates from an ideal solution model for $\delta$ values as low as 0.007. It is characterized experimentally by studying the dependence of nonstoichiometry (or a dependent property there of, such as electrical conductivity)on the partial pressure of oxygen. It can also be studied by extracting the nonstoichiometry dependence of entropy change associated with vacancy formation.  Momentarily disregarding the entropy of gaseous oxygen released upon vacancy formation, the entropy change for the solid phase can be directly evaluated by converting the grand canonical output of MC simulations into canonical quantities. 
\begin{equation}
\langle S \rangle = \frac{1}{T}(\langle E \rangle - \Phi - \sum_i\mu_i\langle n_i\rangle) 
\end{equation}
We  fit a model which includes ideal configurational entropy with a polynomial correction term (to account for non-ideal behavior) to S($\delta$) evaluated per site.
\begin{equation}
\begin{split}
S_{Solid}(\delta) = A\left[ 2\left(\frac{\delta}{2}\right)\ln\left(\frac{\delta}{2}\right) + 2\left(1-\frac{\delta}{2}\right)\ln\left(1-\frac{\delta}{2}\right) \right.\\
\left. + \left(2\delta\right)\ln\left(2\delta \right) + \left(1-2\delta\right)\ln\left(1-2\delta\right)\vphantom{\frac{1}{2}}\right]\\
 +B\delta^3 + C\delta^2 + D\delta + E
  \end{split}
\end{equation}
where $\delta$ is the nonstoichiometry, A,B,C,D and E are constants (for a given T). The solid state entropy S$_{Solid}$($\delta$) per site referenced to S$_{Solid}$($\delta=0$) is plotted in Fig.~\ref{fig:entropydeviation} for T=1480 K. The ideal solution entropy(the term whose coefficient is A)  peaks at $\delta=0.25$, but the actual entropy of the system plateaus much sooner. The pronounced deviation from  ideal solution behavior is apparent from $\delta$ values as low as 0.01. Vibrations clearly increase the entropic advantage to having vacancies, but the strongly non ideal character persists. To compare with experimental work\cite{Panlener1975}, we computed the  entropy of reduction associated with forming an oxygen vacancy in the limit of infinitesimal change in nonstoichiometry.
\begin{equation}\label{eq:reaction2}
\lim_{\Delta\delta\rightarrow 0} \frac{1}{\Delta\delta} \mathrm{CeO}_{2-\delta} \longrightarrow  \frac{1}{\Delta\delta}  \mathrm{CeO}_{2-(\delta+\Delta\delta)} + \frac{1}{2} \mathrm{O}_2
\end{equation}
\begin{eqnarray}
\Delta S_{Total}(\delta)= \Delta S_{Solid}(\delta) + 0.5*S_{O_2}(\delta)\\
S_{O_2}(T) = S_{O_2}^0(T, 1\:atm)  - k_B\ln\left[\frac{p_{O_2}}{1\:atm}\right] \\
\Delta S_{Solid}(\delta) = \frac{1}{\Delta\delta} (S_{CeO_{2-\delta}} - S_{CeO_{2-\delta-\Delta\delta}})
\end{eqnarray}
The solid state entropy is readily available from MC simulations. However, there are difficulties associated with calculating entropy of gaseous oxygen  from first principles, mainly concerned with the definition of a reference state. This can be resolved by using  the standard state entropy data for molecular oxygen from thermochemical tables\cite{nist}.  
$p_{O_2}$ can be obtained using Eqn.~\ref{eqn:gasmu}.
\begin{equation}\label{eqn:gasmu}
\mu_O(T) = \mu_O^{0}(T) + k_BT\ln\left[\frac{p_{O_2}}{1\:atm}\right] 
\end{equation} 
While the chemical potentials of the respective species from MC simulations are  guaranteed to yield the right equilibrium composition and free energies, they are arbitrarily displaced from their true values by an additive constant. This prevents a straightforward application of Equation~\ref{eqn:gasmu}. As a workaround, we used the experimentally published nonstoichiometry vs ln($p_{O_2}$) data\cite{Panlener1975} to fit the chemical potential  from MC and obtain the offset. $\Delta S_{Total}(\delta)$ computed using this approach at 1480 K is illustrated in Fig.~\ref{fig:entropyreduction}.
The strong deviation from ideal solution model suggests that the entropic benefit from vacancies is being offset by some kind of defect interactions. These defect interactions could arise from stable defect complexes or vacancy ordering over short length scales. We look to characterize and quantify the short range order (SRO) by calculating  the thermally averaged pair correlations for the pair clusters in real space. Formally, this idea is embodied in the  Warren-Cowley parameters.
\begin{equation}
\alpha_{lmn}(\delta) = \frac{\left<\sigma_{000}\sigma_{lmn}\right> - \left<\sigma_{000}\right>^2}{1-\left<\sigma_{000}\right>^2} 
\end{equation}
\begin{equation}
\left<\sigma_{000}\sigma_{lmn}\right> = \sum_\sigma P(\sigma,T) \frac{1}{N} \sum_i \sigma_i\sigma_{i+(lmn)}
\end{equation}
For a pair \{(0,0,0), (l,m,n)\} in the  O$^{2-}$/Vac sub-lattice  with vacancy site fraction `s', $\hat{\sigma}_i$ is the occupation variable for site i averaged over all sites equivalent by symmetry, $\left<\hat{\sigma}_{000}\right>^2=(1-2s)^2$ is the MC average of the point cluster correlation  and $\left<\hat{\sigma}_{000}\hat{\sigma}_{lmn}\right>$ is the averaged correlation  function of the pair cluster. P($\sigma$,T) is the probability of  configuration $\sigma$, given by
\begin{equation}
P(\sigma,T) = \frac{1}{Z(\{\mu_i\},T)}\exp\left[-\beta\left(E(\sigma)-\sum_i\mu_ix_i\right)\right] 
\end{equation}
where $Z(\{\mu_i\},T)$ is the semi-grand canonical partition function and \{$\mu_i$\} is the set of chemical potentials. If  vacancies behave ideally (non interacting, randomly distributed), then  $\left<\hat{\sigma}_{000}\hat{\sigma}_{lmn}\right> =\left<\hat{\sigma}_{000}\right>\left<\hat{\sigma}_{lmn}\right> =  (1-2s)^2$ and $\alpha_{lmn}(\delta) = 0 $ for any $\delta$. Clustering of like species is given by $\alpha_{lmn} > 0$ and likewise ordering of unlike species is given by $\alpha_{lmn}< 0$.  
\begin{figure}
\includegraphics[scale=0.46]{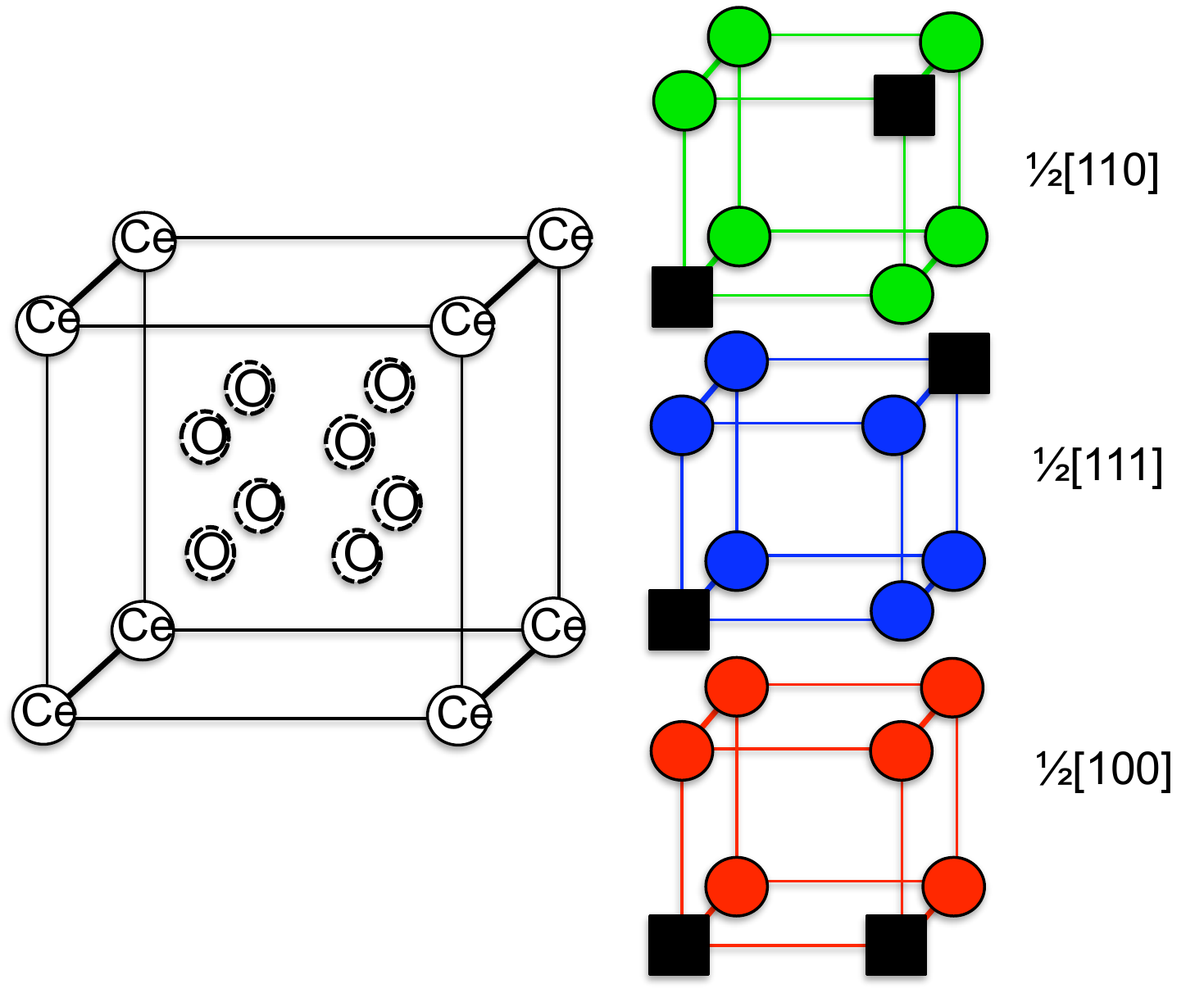}% Here is how to import EPS art
\caption{(Color online) Different vacancy pairs  in the anion sub-lattice of a unit cell of ceria. Black boxes denote vacancies. The color scheme for the oxygen atoms corresponds with that used for plotting SRO parameters of various vacancy pairs in Fig. 8.}
\label{fig:vacancyordering}
\end{figure}
\begin{table}[ht]
\caption{Cluster type, coordinates and diameter of the first five pair clusters in the anion sublattice of ceria. There are 2 distinct 1/2[111] clusters, one within the unit cell(superscript `a') and another extending out of the unit cell, with a face centered Ce atom halfway in between (superscript `b').}
\label{tab:ecis}
\begin{tabular}{|c|c|c|c|}
\hline
Cluster type  & Sites (fractional & Diameter(\AA)  \\
  &  coordinates)  & \\\hline\hline
1/2[100] & (0.25,0.25,0.25) & 2.74  \\
 &  (0.25,0.25,0.75)     & \\\hline
1/2[110] & (0.25,0.25,0.25) & 3.87  \\
&  (-0.25,0.25,0.75)     & \\\hline
1/2[111]$^a$ & (0.75,0.75,0.75) & 4.75 \\
& (0.25,0.25,1.25)      & \\\hline
1/2[111]$^b$ & (0.75,0.75,0.75) & 4.75 \\
&  (0.25,0.25,0.25)     & \\\hline
[100] & (0.75,0.75,0.75) & 5.48\\
& (-0.25,0.75,0.75)      & \\\hline
\end{tabular}
\end{table}

Fig.~\ref{fig:vacancyordering} shows three vacancy pairs in the anion sublattice of a unit cell of ceria. There are 2 distinct 1/2[111] clusters, one within the unit cell and another extending out of the unit cell, with a face centered Ce atom halfway in between. Tab.~\ref{tab:ecis} summarizes the coordinates and diameters of the five smallest pair clusters in the anion sublattice. In  Fig. 8(a), $\alpha(\delta)$ at 1320 K for the five clusters in Table~\ref{tab:ecis} are plotted. Close to stoichiometry, $\alpha_{lmn}=0$ for all (lmn), as would be expected for  non interacting vacancies. However, deviations from zero  become apparent even for slightly off-stoichiometric compositions. In particular, there is strong preference for vacancies to  align along 1/2[110]. The negative correlation along 1/2[100] indicates that two vacancies are  not thermodynamically favored at nearest neighbor sites. The two distinct 1/2[111] pairs have nearly the same value of $\alpha > 0$ and hence are possible directions for short range clustering of vacancies. The effect of temperature is to  favor disorder (Fig. 8(b) for WC parameters at 1791 K), as can be seen from the fact that $\alpha_{lmn} \approx 0$ up to larger nonstoichiometry. At still higher concentrations of vacancies, the $\alpha_{lmn}$ start to deviate from zero and show similar clustering/ordering tendencies along the respective directions. 
\begin{figure}[!ht]
 \subfigure{
\label{fig:wc1}\hspace*{-0.8cm}\includegraphics[scale=0.37]{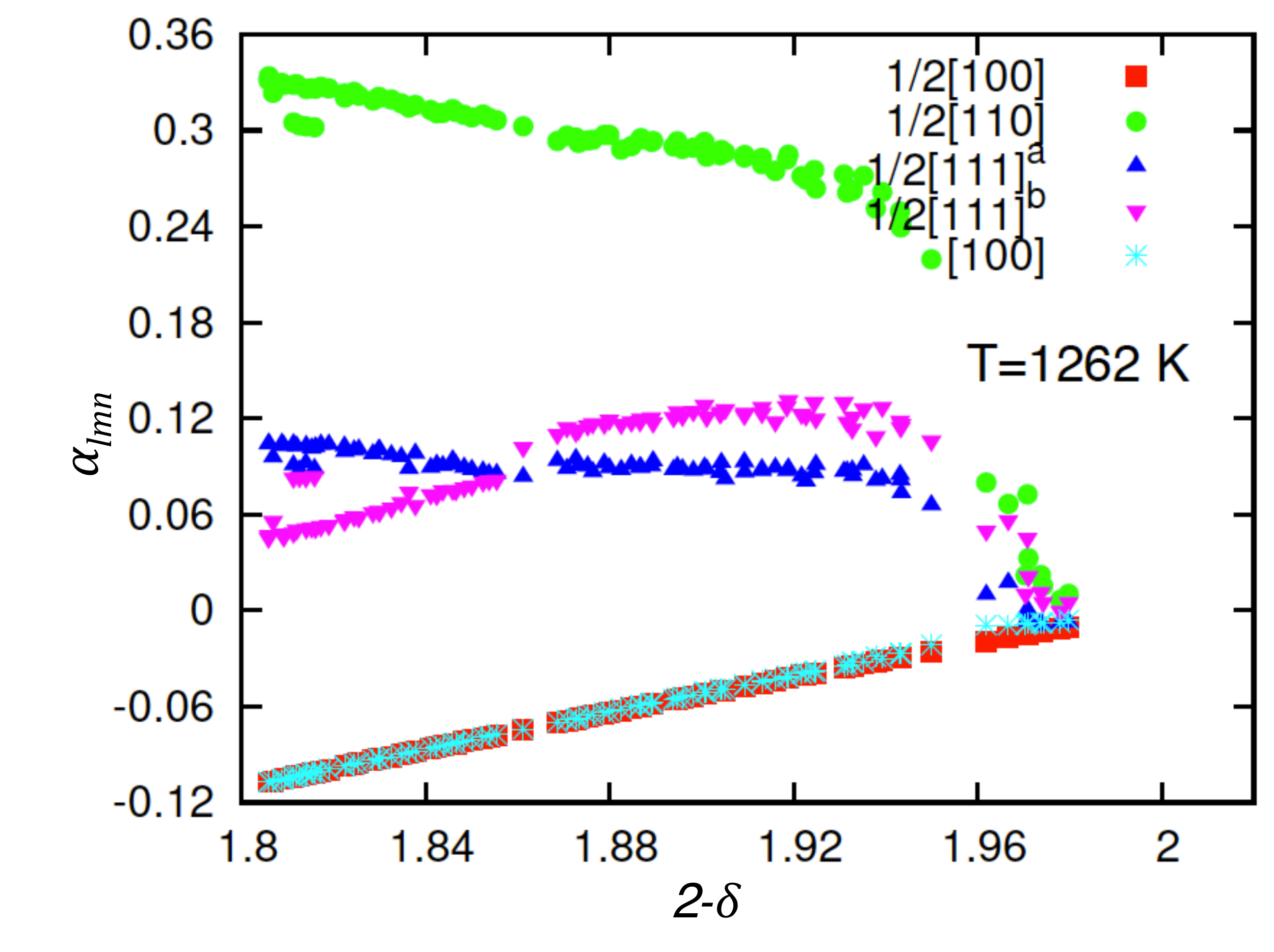}
}
 \subfigure{
\label{fig:wc2}\hspace*{-0.8cm}\includegraphics[scale=0.37]{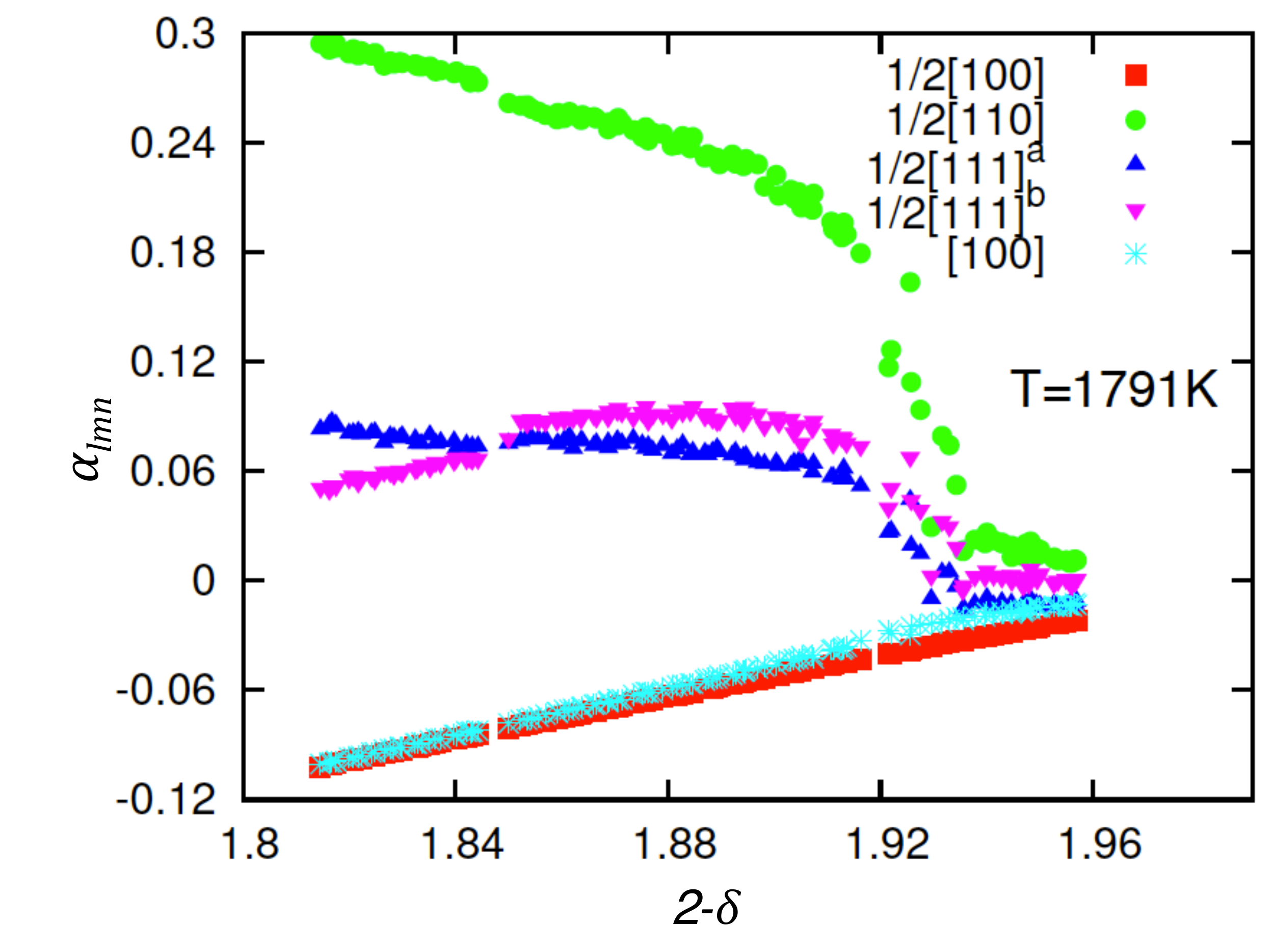}
}
\caption{(Color online) Warren-Cowley short range order parameters ($\alpha_{lmn}$) as a function of nonstoichiometry for various pair clusters in the O$^{2-}$/Vac sublattice at (a) 1262 K and (b) 1791 K. Of note is a strong  clustering tendency along 1/2[110].}
\label{fig:wc}
\end{figure}
 
 There are multiple related system properties associated with vacancy formation in ceria, hence no one rule can govern the choice of dopants. While vacancies are crucial, it is the change in nonstoichiometry between the high and low temperature steps (Eqn.~\ref{eq:reaction}) that ultimately establishes the oxygen uptake and consequently the amount of hydrogen produced. An important implication of this study is that interactions between vacancies (can include dopant-vacancy interactions in doped ceria) markedly diminish the entropic driving force for the thermochemical cycling of ceria. A dopant which tends to increase the dielectric constant of ceria  would screen vacancies from seeing each other and inhibit short range ordering. A computational study combining  accurate first principles calculations of  electronic structure and energetics, together with an efficient cluster expansion based Monte Carlo simulations of configurational disorder, temperature and oxygen chemical potential effects is essential to screen potential dopants for ceria.

\section{Conclusions}\label{sec:conclusions}
We  successfully employed a cluster expansion Hamiltonian based lattice Monte Carlo simulations approach to quantitatively compute the high  temperature thermodynamics of oxygen vacancies in ceria  from first principles. The  ground state energy and electronic structure of nonstoichiometric ceria were obtained from GGA+\textit{U} supercell calculations. The database of structures and  energies was used  to fit a coupled cluster expansion that explicitly accounts for charge state disorder (Ce$^{3+}$/Ce$^{4+}$). Lattice vibrational free energies were calculated from first principles under the harmonic approximation and found to favor formation of vacancies. Vibrational effects were incorporated as a temperature correction to the ECIs. The phase diagram obtained from lattice Monte Carlo simulations was found to exhibit a miscibility gap. The inclusion of vibrations resulted in  quantitative corrections to the composition and temperature range of the miscibility gap, yielding excellent agreement with experiments. The solid state entropy change resulting from vacancy  formation was evaluated and the deviation from ideal solution behavior illustrated through composition dependence of entropy. To further quantify the defect interactions leading to deviations from ideality, Warren Cowley short range order parameters were computed. It was found that there is a strong preference for vacancies to cluster along 1/2[110] and 1/2[111] directions, while the nearest neighbor 1/2[100] sites exhibited ordering behavior. While temperature does disorder the structure, the aforementioned behavior was shown to persist at temperatures as high as 1780 K.

\section*{Acknowledgements}
Work supported by the National Science Foundation under CAREER Grant DMR-1154895 
and by Teragrid/XSEDE computational Resources provided by NCSA and TACC under 
Grant No. DMR050013N. 

\bibliographystyle{apsrev}
\bibliography{allpapers_nourl}

\end{document}